\documentstyle[11pt,fleqn]{article}
\newcommand{\preprint}[1]{\hfill{\sl preprint - #1}\par\bigskip\par}
\def\title{\par\bigskip\begin{center}\LARGE}
\def\endtitle{\end{center}\par\bigskip\par\normalsize}
\def\instit{\begin{center}\it\large}
\def\endinstit{\end{center}\rm\normalsize}
\def\references{\begin{thebibliography}{10}}
\def\endreferences{\end{thebibliography}\end{document}}
\newcommand{\sect}[1]{\section{#1}\renewcommand{\theequation}
        {\mbox{\arabic{section}.\arabic{equation}}}\setcounter{equation}{0}}
\newcommand{\app}[1]{\section{Appendix: #1}\renewcommand{\theequation}
          {\mbox{\Alph{section}.\arabic{equation}}}\setcounter{equation}{0}}
\renewcommand{\author}[1]{\begin{center}\Large #1\end{center}}
\renewcommand{\date}[1]{\par\bigskip\par\sl\hfill #1\par\medskip\par}
\newcommand{\dip}{\smallskip\it Dipartimento di Fisica,
                                Universit\`a di Trento, Italia}
\newcommand{\infn}{\smallskip\it Istituto Nazionale di Fisica Nucleare,\\
                                 Gruppo Collegato di Trento, Italia}
\newcommand{\dinfn}{\dip\\ and\mbox{ }\infn}
\newcommand{\email}[1]{e-mail: \sl #1@itnvax.cineca.it,
                               #1@itncisca.bitnet, itnvax::#1}
\newcommand{\femail}[1]{\footnote{\email{#1}}}
\newcommand{\pacs}[1]{\smallskip\noindent{\sl PACS number(s):
                       \hspace{0.3cm}#1}\par\bigskip}
\newcommand{\babs}{\hrule\par\begin{description}\item{Abstract: }\it}
\newcommand{\eabs}{\par\end{description}\hrule\par\medskip\rm}
\newcommand{\ack}[1]{\par\section*{Acknowledgments} #1}
\newcommand{\R}{\mbox{I$\!$R}}         
\newcommand{\ca}[1]{{\cal #1}}         
\newcommand{\hs}{\qquad\qquad}         
\newcommand{\nn}{\nonumber}            
\newcommand{\ap}{\left.}               
\newcommand{\at}{\left(}               
\newcommand{\aq}{\left[}               
\newcommand{\cp}{\right.}              
\newcommand{\ct}{\right)}              
\newcommand{\cq}{\right]}              
\newcommand{\cg}{\right\}}             
\newcommand{\beq}{\begin{eqnarray}}                    
\newcommand{\eeq}{\end{eqnarray}}                      
\newcommand{\beqn}{\begin{eqnarray}}                   
\newcommand{\eeqn}{\end{eqnarray}}                     
\newcommand{\fr}[2]{\mbox{$\frac{#1}{#2}$}}      
\newcommand{\tr}{\,\mbox{tr}\,}                  
\newcommand{\al}{\alpha}
\newcommand{\be}{\beta}

\newcommand{\de}{\delta}
\newcommand{\ep}{\varepsilon}

\newcommand{\la}{\lambda}

\newcommand{\ph}{\varphi}

\newcommand{\La}{\Lambda}

\begin{document}

\newcommand{\q}{\left(\xi-\fr16\right)}

\newcommand{\hp}{64\pi^2}
\newcommand{\lm}{\log\frac{m^2}{\mu^2}}
\newcommand{\lmz}{\log\frac{M_0^2}{\mu^2}}
\newcommand{\lmu}{\log\frac{M_1^2}{\mu^2}}
\newcommand{\lmd}{\log\frac{M_2^2}{\mu^2}}
\newcommand{\lmt}{\log\frac{M_3^2}{\mu^2}}
\newcommand{\lmq}{\log\frac{M_4^2}{\mu^2}}
\newcommand{\lmc}{\log\frac{M_5^2}{\mu^2}}
\newcommand{\lms}{\log\frac{M_6^2}{\mu^2}}
\newcommand{\lmse}{\log\frac{M_7^2}{\mu^2}}
\newcommand{\lmo}{\log\frac{M_8^2}{\mu^2}}
\newcommand{\lmn}{\log\frac{M_9^2}{\mu^2}}
\newcommand{\lmemz}{\log\frac{M^2}{M_0^2}}
\newcommand{\lmemu}{\log\frac{M^2}{M_1^2}}
\newcommand{\lmemd}{\log\frac{M^2}{M_2^2}}
\newcommand{\lmemt}{\log\frac{M^2}{M_3^2}}
\newcommand{\lmemq}{\log\frac{M^2}{M_4^2}}
\newcommand{\lmemc}{\log\frac{M^2}{M_5^2}}
\newcommand{\lmems}{\log\frac{M^2}{M_6^2}}
\newcommand{\lmemse}{\log\frac{M^2}{M_7^2}}
\newcommand{\lmemo}{\log\frac{M^2}{M_8^2}}
\newcommand{\lmemn}{\log\frac{M^2}{M_9^2}}
\newcommand{\lmem}{\log\frac{M^2}{m^2}}
\newcommand{\lme}{\log\frac{M^2}{\mu^2}}
\newcommand{\clt}{\tilde{\phi}}
\newcommand{\cl}{\hat{\phi}}
\newcommand{\dvol}{dvol({\cal M})}
\newcommand{\inm}{\int_{\cal M}}
\newcommand{\pa}{\partial}
\newcommand{\caln}{{\cal N}}
\newcommand{\calm}{{\cal M}}

\preprint{UTF 293}

\begin{title}
Effective Lagrangian for self-interacting scalar field theories
in curved spacetime
\end{title}

\author{Klaus Kirsten\femail{kirsten}}
\begin{instit}
\dip
\end{instit}
\author{Guido Cognola\femail{cognola} and Luciano Vanzo\femail{vanzo}}
\begin{instit}
\dinfn
\end{instit}

\date{April 1993}

\babs
We consider a self-interacting scalar field theory in a slowly varying
gravitational background field. Using zeta-function regularization and
heat-kernel techniques, we derive the one-loop effective Lagrangian up
to second order in the variation of the background field and up to
quadratic terms in the curvature tensors. Specializing to different
spacetimes of physical interest, the influence of the curvature on the
phase transition is considered.
\eabs


\sect{Introduction}
In new inflationary models \cite{albrechtsteinhardt82,linde82},
the effective cosmological constant is obtained from
an effective potential, which includes quantum corrections to the
classical potential of a scalar field \cite{colemanweinberg73}. This
potential is usually calculated in Minkowski space, but to be fully
consistent, the effective potential or more generally the effective action
must be calculated for more general spacetimes, taking into account
dynamics, geometry and topology of the spacetime itself.
In order to analyze
the influence of these properties of the spacetime on the effective
action in a self-interacting theory,
a variety of methods has been developed in the last years.
In the context of general considerations not refering
to a special spacetime let us mention the quasilocal approximation
scheme for slowly varying background gravitational field at zero
temperature \cite{oconnorhu84} and non-zero temperature
\cite{critchleyhustylianopoulos87}, furthermore the
renormalization-group approach especially elaborated in
\cite{buchbinderodintsovshapiro89,odintsov91,buchbinderodintsovshapiro92}.
In addition there are a lot of
calculations in specific background spacetimes, paying special
attention on the role of constant curvature \cite{oconnorhushen83}, on
topology [10-18],
\nocite{toms82,fordyoshimura79,ford80,toms80,toms80a,denardospallucci80,denardospallucci80a,actor90a,elizalderomeo90a}
on a combination of both
\cite{kennedy81,cognolakirstenzerbini93,bytsenkokirstenodintsov93} and
finally on the anisotropy in different Bianchi type universes [22-29].
\nocite{oconnorhushen85,oconnorhu86,futamase84,huang90,ringwald87,ringwald87a,hartlehu79,berkin92}

In this paper we will use heat-kernel techniques to derive the
one-loop effective Lagrangian for varying background fields up to
second order in the variation of the background and up to quadratic
terms in the curvature tensors
(similar techniques have been applied in
\cite{guven87} and more recently in
\cite{balakrishnantoms92}).
Only under this restrictive condition when the background field
changes very slowly compared to the fluctuation field, it makes sense
to adhere to the effective potential formulation of symmetry breaking
\cite{hu83,chenhu85}. In order to obtain this expansion, we make use
of the partially summed form of the Schwinger-DeWitt asymptotic
development of the effective action introduced by Jack and Parker
\cite{jackparker85}. The nonrenormalized effective Lagrangian is
obtained in Sec.~\ref{s:Ea} and the renormalization
is performed in Sec.~\ref{s:ren}.
Especially we obtain the small background field expansion relevant for
the discussion of a phase transition.

In the following sections, these general results are applied to
several
special cases. In Sec.~\ref{s:symm-space} we treat some maximally
symmetric spaces and the effective potential is found up to quadratic
order in the scalar curvature $R$ and the effect of the curvature on
the phase transition is examined.
In Sec.~\ref{s:taub} we consider the static Taub universe in the limit of
large anisotropy (for the small anisotropy expansion see
\cite{oconnorhushen85}).
The influence of a possible rotation on the phase transition of the
early universe in analyzed using the
example of a G\"odel spacetime in Sec.~\ref{s:goedel}.
Furthermore, in Sec.~\ref{s:bianchi} we easily extract the
results for the most general Bianchi-type I model
(for several limiting cases of this model see
\cite{oconnorhushen85,oconnorhu86,futamase84,hartlehu79,berkin92,husinha88})
deriving on the one side
more detailed results than the ones given up to now
(higher order in the curvature and with the assumption of varying
background field) and deriving on the other side new results
by including a net expansion of the universe together with
nonvanishing shear (generalizing \cite{husinha88,berkin92}).
The conclusions summarize our results. In the appendices we state some
necessary tensor identities and the curvature tensors of the Bianchi-type
I model.

Throughout the paper we will use units in which $\hbar=c=G=1$.

\sect{Effective action}
\label{s:Ea}

The aim of this section is to derive a quasilocal approximation for
the effective action of a self-interacting massive scalar field
coupled to a n-dimensional smooth
background spacetime ${\cal M}$ with Lorentzian metric $g_{\mu \nu}=
diag(-,+,...,+)$ and scalar curvature $R$.
The classical action describing the theory is given by
\beq
S[\clt,g_{\mu\nu}]=-\int_{{\cal M}}\left[\fr12\clt L\clt
+V(\clt)\right]dvol{\cal M}\label{21}
\eeq
where $L=-\Box+m^2 +\xi R$ ($\Box$ is the D'Alembert
operator of the manifold ${\cal M}$) and $V(\clt)$
is a potential describing the
self-interaction of the scalar field and which contains furthermore
local expressions of dimension $n$, involving curvature tensors and
non-quadratic terms in the field, independent up to a total divergence.
These latter terms have in general to be included to ensure the
renormalizability of the theory \cite{toms82}.
The action (\ref{21}) has a minimum at $\clt =\cl$ which satisfies the
classical equation of motion
\beq
L \cl +V'(\cl )=0\label{22}
\eeq
the prime denoting derivative with respect to $\cl$.
Quantum fluctuations $\phi =\clt -\cl$ around the classical background
$\cl$ satisfy an equation of the form (to lowest order in $\phi$)
\beq
A\phi =\left(L+V''(\cl )\right)\phi =0\label{23}
\eeq
In the functional integral perturbative approach, the effective action
is expanded in powers of $\hbar$ as
\beq
\Gamma[\cl]=S[\cl]+\Gamma^{(1)}+\Gamma'\label{24}
\eeq
where $S[\cl]$ is the classical action and the one-loop
contribution $\Gamma^{(1)}$ to the action is given by
\beq
\Gamma^{(1)}=\frac i2\log\det\frac A{\mu^2}
\label{25}
\eeq
The term $\Gamma'$ represents higher loop corrections which we are
not going to discuss. The introduction of the arbitrary mass
parameter $\mu$ is necessary in order to keep the action
dimensionless.

The action (\ref{24}) may be expanded in terms of derivatives of the
background field, that is
\beq
\Gamma[\cl]=\int_{{\cal M}}\left[V(\cl)+\fr12
Z(\cl)\cl_{;\mu}\cl^{;\mu}+...\right]dvol({\cal M})\label{26}
\eeq
The zero derivative term is called the effective potential. In a
static homogeneous spacetime, $\cl $ is a constant field and the
effective action reduces to $\Gamma [\cl ] =vol({\cal M})V(\cl)$.
Under these conditions the concept of the effective potential is well
defined, otherwise one has to work with the effective action
\cite{oconnorhu84}.

As mentioned in the introduction, our first tool will be to compute
the quantities $V(\cl)$ and $Z(\cl)$ in eq.~(\ref{26}) up to
quadratic powers in curvature tensors and up to
second derivatives of curvature terms. The goal to achieve this result
will be the use of zeta-function regularization in combination with
the use of heat-kernel techniques. In the zeta-function regularization
scheme the functional determinant in eq.~(\ref{25}) is defined by
\cite{hawking77,critchleydowker76}
\beq
\Gamma^{(1)}[\cl]=-\fr i2\zeta'_{A/\mu^2}(0)
\label{27}
\eeq
where $\zeta_A (s)$ is the zeta-function associated with the operator
$A$, eq.~(\ref{23}), and the prime denotes differentiation with
respect to $s$. This means with $\lambda_j$
as the eigenvalues of $A$, the zeta-function $\zeta_A(s)$ is defined
by
\beq
\zeta_A (s)&=&\sum_j \lambda_j^{-s}
= \frac {i^s} {\Gamma(s)} \sum_j \int_0^{\infty}dt
\,\,t^{s-1}e^{-i\lambda_j t}\nn\\
&=&\frac {i^s}{\Gamma(s)}
\int_0^{\infty}dt\,\,t^{s-1} \tr K(x,x,t)
\label{28}
\eeq
where the kernel $K(x,x',t)$ satisfies the equation
\beq
i\frac {\partial}{\partial t} K(x,x',t)&=&AK(x,x',t)\nn\\
\lim_{t\to 0}K(x,x',t) &=& |g|^{-\fr 1 2} \de (x,x')\label{29}
\eeq
In order to obtain the derivative expansion, eq.~(\ref{26}), of
the effective action, the following ansatz by Jack and Parker is
suggested \cite{jackparker85}
\beq
K(x,x',t)=-i\frac{\Delta_{VM}(x,x')}{(4\pi it)^{n/2}}
\;\Omega(x,x',t)\;
\exp\{i(\frac{\sigma^2(x,x')}{4t}-tM^2)\}
\label{210}
\eeq
where $\sigma(x,x')$ is the proper arc length along
the geodesic $x'$ to $x$, $\Delta_{VM}(x,x')$ is the Van
Vleck-Morette determinant and
$M^2=m^2+(\xi-1/6)R+V''(\cl)$.
For $t\to 0$ the function $\Omega(x,x',t)$
may be expanded in an asymptotic series
\beq
\Omega (x,x',t) =\sum_{l=0}^{\infty}a_l(x,x') (it)^l\label{211}
\eeq
where the coefficients $a_l$ have to fulfill some recurrence relations.
Using the ansatz (\ref{210}) it has been shown in
ref.~\cite{jackparker85},
that the dependence of $a_l$, $l=1,...,\infty$,
on the field $\cl$ is only through derivatives of the field.
As a result, the use
of eq.~(\ref{211}) for the zeta-function (\ref{28}) leads to the
expansion
\beq
\zeta_A (s) &=& \frac {-i}{(4\pi)^{n/2}\Gamma(s)}
\sum_{l=0}^{\infty}\Gamma\left(s+l-\fr n 2\right)
\int_{{\cal M}}a_l(x,x) M ^{n-2l-2s}dvol({\cal M})\label{212}
\eeq
which represents already the derivative expansion of the
zeta-function $\zeta_A(s)$, which enables us to find expansion
(\ref{26}) without problems, once the coefficients $a_l(x,x)$ are
known.
On the calculation of the Minakshisundaran-Seeley-DeWitt
coefficients there
is a rapidly increasing amount of literature and nowadays the first
four coefficients are known explicitly for manifolds without boundary
\cite{avramidi91,amsterdamskiberkinoconnor89}
and algebraic programs for the computation of arbitrary
coefficients are available \cite{fullingkennedy88}
(for manifold with boundary see [41-46]).
\nocite{bransongilkey90,cognolavanzozerbini90,mcavityosborn91,dettkiwipf92,dowkerschofield90,mossdowker89}

The expansion is sensible if the effective mass $M ^2$ of the
theory is large compared to the magnitude of some typical curvature
radius $|R|$ of the spacetime, $M ^2 \gg |R|$.

Due to the different pole structure of $\Gamma (s+l-n/2)$ around $s=0$
for odd and even $n$, we need to consider these two cases
separately.

Let us first treat odd $n$. Then obviously $\zeta_A (0)=0$ and for the
unrenormalized one-loop effective action, eq.~(\ref{27}), one
easily finds
\beq
\Gamma_o^{(1)}[\cl] = -\frac 1 {2(4\pi)^{n/2}}
\sum_{l=0}^{\infty}\Gamma\left(l-\fr n 2\right)
\int_{{\cal M}}a_l(x,x)M^{n-2l}dvol({\cal M})
\label{213}
\eeq
So for odd dimensions, the one-loop quantum corrections do not depend
on
the arbitrary renormalization scale $\mu$.

The calculation for even $n=2k$ is slightly more difficult. After some
algebra one finds
\beq
\Gamma_e ^{(1)}[\cl] &=&-\frac1{2(4\pi)^k}
\int_{{\cal M}}\left\{\sum_{l=0}^k(-1)^l
\frac{M^{2l}a_{k-l}(x,x)}{l!}\left(C_l-\log\frac{M^2}{\mu^2}
\right)\right.\nn\\
& &\qquad\qquad
\left.+\sum_{l=1}^{\infty}(l-1)!\frac{a_{k+l}(x,x)}{M^{2l}}\right\}
          dvol({\cal M})\label{214}
\eeq
where we introduced $C_l=\sum_{k=1}^l(1/k)$.

Let us stress, that these results, eqs.~(\ref{213}) and
(\ref{214}), are valid for arbitrary dimensions of spacetime and
for arbitrary
self-interaction potential $V(\cl)$. In the next chapter we shall
proceed with the renormalization of the theory and from now on we
shall restrict ourselves to $n=4$ and to a quartic self-interaction,
that is $V(\cl )=\lambda\cl^4/24$.

\sect{Renormalization}
\label{s:ren}

For $n=4$, eq.~(\ref{214}) reduces to
\beq
\Gamma^{(1)}[\cl]&=&\frac 1 {32\pi^2}\inm\left\{
\frac{M^4}2\left(\lme -\frac32\right)+a_2(x,x)\lme\right.\nn\\
& &\qquad\qquad
\left.-\sum_{l=0}^{\infty}\frac{l!a_{3+l}}{M^{2l+1}}\right\}\dvol
\eeq
We recall that our aim is to obtain the effective action up
to second order in the curvature and in the derivative of the scalar
field. Then we will need the coefficients
$a_l$ up to $a_3$ for the operator $A$.
They are given in appendix \ref{a:heat}.
It has to be noted that in $a_3$ (see eq.~(\ref{c1})),
all terms but one ($\fr13(\la\cl\cl_{;\mu}/2M)^2$)
are of higher order with
respect to our requirements.
Then we disregard all those terms and write
\beq
\Gamma^{(1)}[\cl]&=&\frac1{64\pi^2}\inm\left\{
M^4\at\lme-\frac32\ct+2a_2\lme
-\frac{\la^2\cl^2\cl_{;\mu}\cl^{;\mu}}{6M^2}
\cg\dvol\;\;+\dots
\label{31}
\eeq
The first term in the integral, eq.~(\ref{31}), obviously
corresponds to the Coleman-Weinberg result
in flat space \cite{colemanweinberg73}.

The quantum correction $\Gamma ^{(1)}[\cl]$ depends on the arbitrary
renormalization scale $\mu$, this dependence being removed by the
renormalization procedure which we are going to describe now.
As is well known by now, one is forced to take into consideration the most
general quadratic gravitational Lagrangian \cite{utiyamawitt62} and so
one has to consider the classical Lagrangian
\beq
L_{cl}&=&\eta\Box\cl^2
-\fr 1 2 \cl \Box \cl +\La +\fr 1 {24} \la \cl ^4
+\fr 1 2 m^2 \cl ^2 +\fr 1 2 \xi R \cl^2 \nn\\
&&+\ep_0 R+\fr 1 2 \ep_1 R^2 +\ep_2 C+\ep_3 G +\ep_4 \Box R
\label{32}
\eeq
with the corresponding counterterm contributions
\beq
\de L_{cl}&=&\de\eta\Box\cl^2
+\de\La+\fr 1 {24} \de\la \cl ^4
+\fr 1 2 \de m^2 \cl ^2 +\fr 1 2 \de\xi R \cl^2 \nn\\
&&+\de\ep_0 R+\fr 1 2 \de\ep_1 R^2 +\de\ep_2 C
+\de\ep_3 G +\de\ep_4 \Box R
\label{33}
\eeq
necessary to renormalize all coupling constants.
Of course $\La\sim 0$ and $\eta=1/4$.
By $C$ and $G$ we indicate respectively the square of the Weyl
tensor and the Gauss-Bonnet density. They read
\beq
C&=&R_{\mu\nu\rho\sigma}R^{\mu\nu\rho\sigma}
-2R_{\mu\nu}R^{\mu\nu}+\fr13R^2\\
G&=&R_{\mu\nu\rho\sigma}R^{\mu\nu\rho\sigma}
-4R_{\mu\nu}R^{\mu\nu}+R^2
\eeq

The renormalization conditions are given by \cite{oconnorhu84,berkin92}
\beq
\La &=&L\left|_{\cl =\varphi_0,R=0}\right.\nn\\
\la &=&\frac{\partial^4 L}{\partial \cl^4}\left|_{\cl =\varphi
      _1,R=0}\right.\nn\\
m^2 &=&\frac{\partial ^2 L}{\partial \cl^2}\left|_{\cl =0,
          R=0}\right.\nn\\
\xi &=&\frac{\pa L}{\pa R \pa \cl^2} \left|_{\cl =\varphi_3,
               R=R_3}\right.\nn\\
\ep_0 &=&\frac{\pa L}{\pa R}\left|_{\cl =0, R=0}\right.\label{34}\\
\ep_ 1&=& \frac{\pa ^2 L}{\pa R^2}\left|_{\cl =0, R=R_5}\right.\nn\\
\ep_2 &=& \frac{\pa L}{\pa C}\left|_{\cl =0,R=R_6}\right.\nn\\
\ep_3 &=&  \frac{\pa L}{\pa G}\left|_{\cl =0,R=R_7}\right.\nn\\
\ep_4 &=&  \frac{\pa L}{\pa \Box R}\left|_{\cl =0,R=R_8}\right.\nn\\
\eta &=&  \frac{\pa L}{\pa \Box \cl ^2}\left|_{\cl =\varphi_9,R=0}\right.
      \nn
\eeq

The conditions (\ref{34}) determine the counterterms to be
\beq
\hp \de \La &=&  \la m^2 \varphi_0^2\lm
  -\frac{m^2\varphi_0^2}{2}
  -\frac{\la\varphi_0^4}{24}
  -\la m^2\varphi_0^4
  +\frac{\la\varphi_0^4}{4}\lmu\nn\\
  &&-M_0^4\lmz+\frac{3M_0^4}{2}
  -\frac{\la^4\varphi_0^4\varphi_1^4}{12M_1^4}
  +\frac{\la^3\varphi_0^4\varphi_1^2}{2M_1^2}
\nn\\
\hp \de \la &=& -6\la ^2\lmu
  +\frac{2\la^4\varphi_1^4}{M_1^4}
  -\frac{12\varphi_1^2\la^3}{M_1^2}\nn\\
\hp \de m^2 &=& 2\la m^2\left(1-\lm\right)\nn\\
\hp \de \xi &=& -2\la \lmt \q
    -\frac{2\la^2\q\varphi_3^2}{m_3^2}\nn\\
\hp \de \ep_0 &=& 2m^2 \q \left(1-\lm\right)
     \label{35}\\
\hp \de \ep_1 &=& -2\q^2\lmc \nn\\
\hp \de \ep_2 &=& -\fr1 {60} \lms\nn\\
\hp \de \ep_3 &=& \fr1 {180} \lmse\nn\\
\hp \de \ep_4 &=& \fr1 3 \left(\xi-\fr1 5\right)\lmo\nn\\
\hp\de\eta &=& \fr{\la}{6}\lmn\nn
\eeq
where we introduced $M_i^2=m^2+\q R_i+\frac{\lambda} 2 \varphi_i^2$.
For the sake of generality, we choosed different values
$\ph_i,R_i$ for the definition of the physical coupling constants.
This is due to the fact that in general they are measured at different
scales, the behaviour with respect to a change of scale being
determined by the renormalization group equations.

After some calculations one finds the renormalized
effective Lagrangian in the form
\beq
\lefteqn{\hp L_r^{(1)}=\frac{C}{60}\lmems-\frac{G}{180}\lmemse
-\left(\xi-\fr15\right)\frac{\Box R}{3}\lmemo}\nn\\
&&-32\pi^2m^2\varphi_0^2-\frac{8\pi^2\la\varphi_0^4}{3}+m^4\lmemz
     +\la m^2\varphi_0^2\left(\log\frac{m^2}{M_0^2}+\frac 1 2\right)
\nn\\
& &+2m^2\q R\left(\lmem-\frac12\right)
      +\q^2R^2\left(\lmemc-\frac3 2 \right)\nn\\
& &-\frac{\la^2\varphi_0^4}4
      \left[\log\frac{M_0^2}{M_1^2}-\frac32
      -\frac{4(M_1^2-m^2)(2M_1^2+m^2)}{3M_1^4}\right]
\label{36}\\
& &+\left\{\q R\left[\lmemt-\frac32-\frac{\la\varphi_3^2}{M_3^2}\right]
         +m^2\left[\lmem-\frac1 2\right]\right\}\la\cl^2\nn\\
& &+\left\{\left(\lmemu-\frac{25}6\right)
    +\frac{4m^2(m^2+M_1^2)}{3M_1^4}\right\}\frac{\la^2\cl^4}4
-\frac{\la\Box\cl^2}{6}\lmemn
-\frac{\la^2\cl^2\cl_{;\mu}\cl^{;\mu}}{6M^2}\nn
\eeq

For the discussion of the phase transition of the system, the expansion
of $L_r^{(1)}$ for small values of the
background field is of relevance. It is easily found to be
\beq
L_r^{(1)}&=&\La_{eff}+\frac{\la\cl^2}{64\pi^2}
   \left\{m^2\log\at 1+\q\frac{R}{m^2}\ct
-\frac{\la\cl_{;\mu}\cl^{;\mu}}{6[m^2+\q R]}
    \right.\nn\\
& &+R\q\left[\log\frac{m^2+\q R}{M_3^2}
  -\frac{\lambda\varphi_3^2}{M_3^2}-1\right]\label{37}\\
& &\left.-\frac{\lambda}{12(m^2+\q R)}\aq
\Box\cl^2+2\left(\xi-\fr15\right)\Box R
-\frac{C}{10}+\frac{G}{30}\cq\right\}
+\ca O(\cl^4)\nn
\eeq
where $\Lambda_{eff}$ (the cosmological constant) represents a
complicated expression not depending on the background field $\cl$.
Realizing that we are consistently working only in the small curvature
and slowly varying background field approximation,
the latter equation is equivalent to
\beq
L_r^{(1)}&=&\Lambda_{eff}+\frac{\lambda\cl^2 }{64\pi^2}
\left\{\q R\left[\log\frac{m^2}{M_3^2}
-\frac{\lambda\varphi_3^2}{M_3^2}\right]\right.\nn\\
& &\qquad+\q^2 \frac{R^2}{2m^2}+\frac{C}{120m^2}
-\frac{G}{360m^2}\nn\\
& &\left.\qquad-\frac{\la\cl_{;\mu}\cl^{;\mu}}{6m^2}
-\frac{\la\Box\cl^2}{12m^2}
-\left(\xi-\fr1 5\right)\frac{\Box R}{6m^2}\right\}
+\mbox{higher order terms}
\label{38}
\eeq
{}From this, the known result \cite{ilio75-47-165}
\beq
Z(\cl)=\frac{\la^2\cl^2}{192\pi^2m^2}\label{derivative}
\eeq
immediately follows.

Let us first restrict to static homogeneous spacetimes, so that $\cl$
is a constant field. Restricting to the linear curvature approximation
it is seen, that for $R<0$ and $\xi < 1/6$ the one-loop term will help
to break symmetry, whereas for $\xi > 1/6$ the quantum contribution
acts as a positive mass and helps to restore symmetry. For $R>0$ the
conclusions are obviously reversed. This behaviour has already been
found for several examples (see
\cite{chimentojakubipullin89,berkin92,cognolakirstenzerbini93}),
here it is seen to be valid for any smooth manifold. For non-smooth
manifolds such as orbifolds, conical singularities lead to additional
contributions in eq.~(\ref{c1}) and eq.~(\ref{38})
remains no longer valid (for a recent example see
\cite{bytsenkokirstenodintsov93}).

The influence of the higher order terms as well as
the derivative terms in eq.~(\ref{38}) can in
general not be stated. But in the next sections we will, starting from
the general results (\ref{36})-(\ref{38}), analyze several theories
and so determine in more detail these contributions.

\sect{Self-interacting $\phi ^4$ theory in maximally symmetric
spaces}
\label{s:symm-space}

Let us first consider maximally symmetric spaces and direct products
of them. More explicitly we treat the 4-dimensional manifolds
$\ca M^4,\R\times\ca{M}^3,S^1\times\ca{M}^3,\ca{M}^2\times\ca M^2$,
with $\ca M^i=\tilde\ca{M}^i/\Gamma$,
where $\tilde\ca{M}^i$, which is equal to $\R^i$, $S^i$ or $H^i$
($S^i$ and $H^i$ being the $n$-sphere and the $n$-dimensional
Lobachevsky space respectively), is the covering manifold and
$\Gamma$ is a discrete group of isometries of $\tilde\ca M^i$,
which acts freely and without fixed points.
In appendix \ref{a:symm-space} we summarized
useful formulas about the geometric tensors in these kind of
manifolds.

In the present section the metric is taken to be Euclidean.

For the given examples we may assume a constant background field $\cl$
and so the relevant quantity is the effective potential
$V(\cl)=vol^{-1}({\cal M})\Gamma[\cl]$.
We will concentrate on the small
$\cl$-expansion, eq.~(\ref{38}). The quantum corrections to the
mass of the field are then defined by
\beq
L_r^{(1)}=\Lambda_{eff} +\fr 1 2 m_T^2 \cl ^2 +{\cal O}(\cl^4)
\label{41}
\eeq
and using eq.~(\ref{38}) it is given by
\beq
m_T^2&=&\frac{\lambda}{32\pi^2}\left\{\q R
\left[\log\frac{m^2}{M_3^2}
-\frac{\lambda\varphi_3^2}{M_3^2}\right]\right.\nn\\
& &\left.\qquad+\q^2\frac{R^2}{2m^2}+\frac{C}{120m^2}
-\frac{G}{360m^2}\right\}\label{42}
\eeq
Depending on the manifold $\tilde{M}_i$, the linear curvature helps to
break or to restore symmetry as described below eq.~(\ref{derivative}).

Let us now consider the influence of the higher order terms.
For $M_4$ we have $C=0$ and $G =\frac 1 6 R^2$, see
(\ref{a3}), so
\beq
m_T^2&=&\frac{\lambda}{32\pi^2}
\left\{\q R\left[\log\frac{m^2}{M_3^2}
-\frac{\lambda\varphi_3^2}{M_3^2}\right]\right.\nn\\
& &\left.\hs+\left[\left(\xi-\frac16\right)^2
-\frac1{1080}\right]\frac{R^2}{2m^2}\right\}\label{43}
\eeq
For $6\xi\in\left[(1-\sqrt{1/30}),(1+\sqrt{1/30})\right]$ the quadratic
terms help to break symmetry, otherwise they help to restore it.

In the case $\R \times\ca M^3$ we find $C=0$, $G=0$, so as a result
\beq
m_T^2&=&\frac{\lambda}{32\pi^2}\left\{\q
R\left[\log\frac{m^2}{M_3^2}
-\frac{\lambda\varphi_3^2}{M_3^2}\right]
+\q^2\frac{R^2}{2m^2}\right\}\label{44}
\eeq
For these examples the $R^2$-contributions are always positive and help
to restore symmetry.

Finally, for $\ca M=\ca M^2\times\ca N^2$, (\ref{a7}) and (\ref{a8}) yield
$C=\frac 2 3 R(\ca M^2)R(\ca N^2)$ and
$G=2R(\ca M^2)R(\ca N^2)$, which also lead
to eq.~(\ref{44}).

As is well known and as is seen also in these examples, in the small
curvature limit the topology has no visible influence on the presented
results.

This concludes the examples of spaces of constant curvature, and we
will now consider the static Taub universe.

\sect{Self-interacting $\phi^4$ theory in a static Taub universe}
\label{s:taub}

The metric of a diagonal mixmaster universe is given by
\beq
ds^2 =-dt^2+\sum_{a=1}^3l_a^2(\sigma^a)^2
\label{51}
\eeq
where $\sigma ^a$ are the basis one-forms
\beq
\sigma^1&=&\phantom{-}\cos\psi\,\,d\theta +\sin\psi\,\,
\sin\theta\,\,d\phi\nn\\
\sigma^2&=&-\sin\psi\,\,d\theta
+\cos\phi\,\,\sin\theta\,\,d\phi\label{52}\\
\sigma^3&=&\phantom{-}\cos\theta\,\,d\phi+d\psi\nn
\eeq
The $l_a$'s are the three principal curvature radii of the homogeneous
space and are constants for a static universe. The cases when any two
of the $l_a$'s are equal are the Taub universes. The case when all
three $l_a$'s are equal is the closed Friedmann-Robertson-Walker
universe.

The Taub universe has been first considered in \cite{oconnorhushen85}
with the aim of analyzing the effect of curvature anisotropy on
symmetry breaking. This was a continuation of investigations into the
symmetry behaviour of a self-interacting field in curved spacetime
using the Einstein universe as an example. The Taub universe was then
treated as a perturbative expansion around the Einstein universe in
powers of the anisotropy $\alpha=(l_1^2/l_3^2)-1$, thus the results
obtained are valid for small anisotropy.

In terms of the anisotropy the scalar curvature $R$ is given by
\beq
R&=&\frac{4l_1^2-l_3^2}{l_1^4}=
\frac{(3+4\alpha)}{2l_1^2(1+\alpha)}
\label{53}\\
&=&\frac3{2l_1^2} \left(1+\frac{\alpha} 3
+{\cal O}(\alpha^2)\right)\nn
\eeq
where the expansion for small values of $\alpha$ makes explicit the
expansion around the Einstein universe.

As already mentioned, our expansion is consistent for small values of
the curvature (compared to the mass $m$ of the field).
Defining $l_1=\sqrt{N+1}\;l_3$, that is $\alpha=N$, we obtain
\beq
R=\frac{4N+3}{2(N+1)^2l_3^2}\label{54}
\eeq
Thus for fixed values of $l_3$, we see that our expansion is sensible
for large values of $N$, which means large anisotropy $\alpha$, so
we deal with the
range not considered in \cite{oconnorhushen85}. The relevant
quantities for the mass $m_T^2$ are \cite{oconnorhushen85}
\beq
G=0, \hs  C&=&\frac{4N^2}{3(N+1)^4l_3^4},
\label{56}
\eeq
So we find
\beq
m_T^2&=&\frac{\lambda}{32\pi^2}\left\{
\q\frac{4N+3}{2(N+1)l_3^2}\left[\log\frac{m^2}{M_3^2}
-\frac{\lambda\varphi_3^2}{M_3^2}\right]\right.\nn\\
& &\left.+\frac 1 {2m^2(N+1)^4l_3^4}
\left[\q^2\frac{(4N+3)^2}{4}+\frac{N^2}{45}\right]\right\}\label{57}
\eeq
where once more higher order terms in the curvature help to restore
the symmetry.

\sect{Self-interacting $\phi ^4$ theory in a G\"odel spacetime}
\label{s:goedel}

Our next example will be the G\"odel spacetime in order to consider
the question of how a possible rotation of the early universe
influences the phase transition.
In the G\"odel case, as in the static metrics, the classical field can
be chosen to be constant.
This model has been recently
considered for a massless scalar field theory
\cite{chimentojakubipullin89} making use of the operative continuation
renormalisation method \cite{chimentojakubi89}. We generalize these
results to the massive case.

The G\"odel metric is defined by
\beq
ds^2=-dt^2 +dx^2 -\fr12\exp(2\sqrt{2}\omega x)dy^2
-2\exp(\sqrt{2} \omega x)dt\,\,dy+dz^2
\label{71}
\eeq
where $\omega$ is the so called vorticity, a measure of the constant
rotation of the matter flow. As a function of the vorticity, the
relevant geometric tensors read
\beq
R=-2\omega^2,\qquad G=0,
\qquad C=\fr{16}3\omega^4,
\label{72}
\eeq
which leads to the mass
\beq
m_T^2&=&\frac{\lambda}{16\pi ^2}\left\{-\q\omega ^2\left[
\log\frac{m^2}{M_3^2}-\frac{\lambda \varphi_3^2}{M_3^2}\right]
\right.\nn\\
& &\left.\qquad+\frac{\omega^4}{m^2}
\left[\at\xi-\frac16\ct^2+\frac1{45}\right]\right\}
\label{73}
\eeq
Also in this case, the corrections to the linear curvature term help
to restore the symmetry.

\sect{Self-interacting $\phi ^4$ theory in a Bianchi type-I
universe}
\label{s:bianchi}

Let us now consider the Coleman-Weinberg symmetry breaking mechanism
in a Bianchi type-I universe with the metric
\beq
ds^2 =-dt^2 +\alpha(t)\sum_{i=1}^3e^{2\beta_i(t)}(dx^i)^2\label{61}
\eeq
with the $\beta_i$ taken to be traceless, $\sum_{i=1}^3\beta_i=0$, so
that $g=\alpha^3(t)$.
$\alpha(t)$ is a positive parameter called $a^2$
in ref.~\cite{berkin92}, where
this model was very recently considered with the
aim of investigating the influence of the shear on the symmetry
breaking mechanism. In this reference $\alpha(t)=1$ was considered and
for small anisotropy $\beta_i(t)$
the influence of curvature up to linear orders in $R$ was found.
The expansion of ref.~\cite{berkin92} is consistent for small scalar
curvature $R$, which is identical to say for small shear.
But this is exactly the range our results are valid. So using
our general expansion, we will extend the results of \cite{berkin92} to
nonconstant $\alpha(t)$ and arbitrary anisotropy up
to quadratic orders in the curvature, that is
equivalent to fourth order derivative terms in the scale factors
$\alpha(t)$ and $\beta_i(t)$.

The scalar curvature of the Bianchi type-I model (\ref{61}) together
with
the Gauss-Bonnet density and the Weyl tensor are stated in appendix
\ref{a:bianchi}
(see eqs.~(\ref{b1})-(\ref{b3})). It is seen, that small
curvature expansion means slowly varying scale factor $\alpha(t)$ and
anisotropy $\beta_i(t)$, with no restriction on the magnitude of
$\alpha(t)$ and $\beta_i(t)$ themselves.

For nonconstant $\alpha(t)$ for the sake of simplicity, we will only state
the linear curvature term. However, in principle also quadratic order
terms in the curvature are easily
obtained from the tensors given in appendix \ref{a:bianchi}, but this
result would be difficult to survey and we will give it only for
constant $\alpha$.

Concentrating on the effective
potential part, the mass for nonconstant $\alpha(t)$ in the linear
curvature approximation reads
\beq
m_T^2&=&\frac{\lambda}{32\pi^2}
\q\left[\frac{3\alpha''}{\alpha}+Q_1\right]
\left[\log\frac{m^2}{M_3^2}
-\frac{\lambda\varphi_3^2}{M_3^2}\right]
\label{62}
\eeq
In eq.~(\ref{62}), $Q_1=\sum_{i=1}^3(\beta_i')^2$ is a measure of
the shear and the prime denotes differentiation with respect to $t$.
Comparing with the result of Berkin~\cite{berkin92},
there are of course additional
contributions due to the nonconstant scale factor $\alpha$, but the
general feature in terms of the scalar curvature as stated in
eq.~(\ref{38}) remains true.

For constant $\alpha$ up to fourth order derivative terms the result reads
\beq
m_T^2&=&\frac{\lambda}{32\pi^2}\left\{
Q_1\q\left[\log\frac{m^2}{M_3^2}
-\frac{\lambda\varphi_3^2}{M_3^2}\right]\cp\nn\\
&&\ap+\frac{Q_1^2}{2m^2}\aq\left(\xi-\frac16\right)^2+\frac1{45}\cq
+\frac{Q_2}{60m^2}+\frac{2}{45m^2}
\frac{d}{dt}\at\beta_1'\beta_2'\beta_3'\ct
\right\}
\eeq
where $Q_2==\sum_{i=1}^3(\beta_i'')^2$ was set and
explicit use of the $\beta_i$ to be traceless was made.
Depending on the $\beta_i$'s, the fourth order derivative terms may
help to restore or to break symmetry.
When the $\beta_i'$'s are constants, the fourth order derivative terms
always help to restore symmetry.

Finally, let us concentrate on $\beta_i=0$, $i=1,2,3$
(This is the case of the spatially flat Robertson-Walker
metric; see \cite{husinha88}).
For this case we find
\beq
m_T^2&=&\frac{\lambda}{32\pi^2}
\left\{3\q\frac{\alpha''}{\alpha}
\left[\log\frac{m^2}{M_3^2}
-\frac{\lambda\varphi_3^2}{M_3^2}\right]\right.\nn\\
&&\ap+\frac{9}{2m^2}\q^2\at\frac{\alpha''}{\alpha}\ct^2
-\frac{1}{240m^2}\at\frac{\alpha'}{\alpha}\ct^2
\aq\frac{2\alpha''}{\alpha}-\at\frac{\alpha'}{\alpha}\ct^2\cq
\right\}
\label{64}
\eeq
Once more, depending on the scaling function $\alpha$, higher curvature
terms may help to restore or to break symmetry.
More precisely, if $\alpha''$ is negative or greater than
$(\alpha')^2/15\alpha$, the symmetry may be restored
with the help of higher curvature terms.

\sect{Conclusions}

In new inflationary models, the effective cosmological constant is
obtained from an effective potential, which includes quantum
corrections to the classical potential of a scalar field
\cite{colemanweinberg73}. This potential is usually calculated in
Minkowki spacetime, whereas to be fully consistent the effective
potential must be calculated for more general spacetimes. For that
reason an intensive research has been dedicated to the analysis of the
one-loop effective potential of a self-interacting scalar field in
curved spacetime and in spacetimes with nontrivial topology [4-29].

In this paper we provided a simple approach for the calculation of the
derivative expansion of the effective Lagrangian in the background
field, the expansion being consistent if the effective mass $M^2=m^2
+\left(\xi -\frac 1 6\right)R+\frac{\la} 2 \cl ^2$ of the theory is
large compared to a typical magnitude of the curvature. The result has
been used for several examples of physical
interest to analyze the influence of
the gravitational field on the phase transition.

In the linear curvature approximation it is seen, that for $R<0$ and
$\xi <\frac 1 6$ the one-loop term will help to break symmetry,
whereas for $\xi >\frac 1 6 $ the quantum contribution acts as a
positive mass and help to restore symmetry. The conclusions are
reversed if we choose $R>0$. The influence of the higher curvature
terms depends on the spacetime under consideration and is stated
explicitly in the respective sections.

Using the provided approach it was very simple to extract the relevant
information for several spacetimes, thus extending
recent results.

\ack{K. Kirsten is grateful to the Theoretical Group of the
Department of Physics of the University of Trento for the
kind hospitality.}

\appendix
\app{heat coefficients}
\label{a:heat}

As mentioned in Sec.~\ref{s:Ea}, in this appendix we report on some results
concerning the heat-kernel expansion of a second order elliptic
differential operator. The operator of interest in the given
considerations is $A=-\Box+X(x)$, defined on a smooth n-dimensional
Riemannian manifold without boundary.

Using the ansatz of Jack and Parker, see
eq.~(\ref{210}), the explicitly needed coefficients of
expansion (\ref{211}) read
(for details see ref.~\cite{jackparker85})
\beq
a_0(x,x)&=&1\nn\\
a_1(x,x)&=&0\nn\\
a_2(x,x)&=&-\frac16\Box(X-R/6)
+\frac1{180}\left(\Box R+
R_{\mu\nu\rho\sigma}R^{\mu\nu\rho\sigma}
-R_{\mu\nu}R^{\mu\nu}\right)\label{c1}\\
a_3(x,x)&=&\fr1{12}X^{;\mu}X_{;\mu}
-\fr1{60}\Box^2X
+\fr1{90}R^{\mu\nu}X_{;\mu\nu}
-\fr1{30}R^{;\mu}X_{;\mu}\nn\\
&&+\frac{1}{7!}\at 18\Box^2R+17R_{;\mu}R^{;\mu}
-2R_{\mu\nu;\rho}R^{\mu\nu;\rho}
-4R_{\mu\nu;\rho}R^{\mu\rho;\nu}\cp\nn\\
&&+9R_{\mu\nu\rho\sigma;\tau}R^{\mu\nu\rho\sigma;\tau}
-8R_{\mu\nu}\Box R^{\mu\nu}
+24R_{\mu\nu}R^{\mu\rho;\nu}{}_{\rho}
+12R_{\mu\nu\rho\sigma}\Box R^{\mu\nu\rho\sigma}\nn\\
&&+\fr{208}9R_{\mu\nu}R^{\mu\rho}R^\nu_\rho
+\fr{64}3R_{\mu\nu}R_{\rho\sigma}R^{\mu\rho\nu\sigma}
+\fr{16}3R_{\mu\nu}R^\mu{}_{\rho\sigma\tau}R^{\nu\rho\sigma\tau}\nn\\
&&\ap+\fr{44}9R_{\mu\nu\rho\sigma}R^{\mu\nu\al\be}R^{\rho\sigma}{}_{\al\be}
+\fr{80}9R_{\mu\nu\rho\sigma}R^{\mu\al\rho\be}R^\nu{}_\al{}^\sigma{}_\be
\ct\nn
\eeq
The $a_2$ coefficient fixes the counterterms in the renormalization
procedure, see eq.~(\ref{32}), while only the first term in $a_3$ is
relevant in the approximation in which we are working.

\app{manifolds with constant curvature}
\label{a:symm-space}

In this appendix we want to summarize some tensor identities used in
section \ref{s:symm-space}.

For maximally symmetric $n$-dimensional space ${\cal N}$ one has
\beq
R_{\mu\nu\rho\sigma}({\cal N})&=&
\frac{R(\caln)}{n(n-1)}(g_{\mu\rho}g_{\nu\sigma}
-g_{\mu\sigma}g_{\nu\rho})
\label{a1}\\
R_{\mu\nu}&=&\frac R n g_{\mu\nu}
\eeq
which leads to
\beq
C(\caln)=0,\hs\hs
G(\caln )=\frac{(n-2)(n-3)}{n(n-1)}R^2(\caln)
\label{a3}
\eeq
For constant curvature spaces $\calm$ which are direct products of
maximally symmetric spaces, $\calm =\calm_1\times \calm_2$, the
following identities are useful \cite{bransongilkey90}
\beq
R^2(\calm)&=&R^2(\calm_1)+R^2(\calm_2)
+2R(\calm_1)R(\calm_2)
\label{a4}\\
R_{\mu\nu}(\calm )R^{\mu\nu}(\calm)&=& R_{\mu\nu}(\calm_1)
R^{\mu\nu}(\calm_1)+R_{\mu\nu}(\calm_2)
R^{\mu\nu}(\calm_2)
\label{a5}\\
R_{\mu\nu\rho\sigma}(\calm)R^{\mu\nu\rho\sigma}(\calm )&=&
R_{\mu\nu\rho\sigma}(\calm_1)R^{\mu\nu\rho\sigma}(\calm_1)+
R_{\mu\nu\rho\sigma}(\calm_2)R^{\mu\nu\rho\sigma}(\calm_2)
\label{a6}
\eeq
Using (\ref{a4})-(\ref{a6}) it is easy to arrive at
\beq
C(\calm )=C (\calm_1)+C(\calm_2)+\frac23R(\calm_1)
R(\calm_2)
\label{a7}
\eeq
\beq
G(\calm)=G(\calm_1)+G(\calm_2)+2R(\calm 1)R(\calm_2)
\label{a8}
\eeq
which are the relevant results used in Sec.~\ref{s:symm-space}.

\app{some invariants for Bianchi I universe}
\label{a:bianchi}

Here we give the relevant tensors,
for our considerations,
up to fourth order derivative in the scale factor $\alpha(t)$ and the
anisotropy $\beta_i(t)$
of the Bianchi type-I model of Sec.~\ref{s:bianchi}.

First, for the scalar curvature one has
\beq
R=\frac{3\alpha''}{\alpha}+Q_1
\label{b1}
\eeq
For the Gauss-Bonnet density we find
\beq
G&=&\frac32\at\frac{\alpha'}{\alpha}\ct^2\aq\frac{2\alpha''}{\alpha}
-\at\frac{\alpha'}{\alpha}\ct^2\cq
-Q_1\aq\frac{2\alpha''}{\alpha}
+\at\frac{\alpha'}{\alpha}\ct^2\cq\nn\\
&&+\frac{2\alpha'}{\alpha}\at 6\be_1'\be_2'\be_3'-Q_1'\ct
+8\frac{d}{dt}\at\be_1'\be_2'\be_3'\ct
\label{b2}
\eeq
and the square of the Weyl-tensor reads
\beq
C&=&\frac{Q_1}{2}\at\frac{\alpha'}{\alpha}\ct^2
+\frac{4Q_1^2}{3}+2Q_2\nn\\
&&+\frac{\alpha'}{\alpha}\at 12\be_1'\be_2'\be_3'+Q_1'\ct
+8\frac{d}{dt}\at\be_1'\be_2'\be_3'\ct
\label{b3}
\eeq
These results are used in Sec.~\ref{s:bianchi} to give the
quantum corrections to the mass of the field up to quadratic order in
the curvature.

\end{document}